**Full Title:**

# Orthogonal factorial designs for trials of therapist-delivered interventions: Randomising intervention-therapist combinations to patients.

**Short Title:**

Factorial designs for trials of therapist-delivered interventions

**Authors:**

Rebecca EA Walwyn (University of Leeds)

Rosemary A Bailey (University of St Andrews)

Arpan Singh (University of Leeds)

Neil Corrigan (University of Leeds)

Steven G Gilmour (King's College London)

**Contact Information for Corresponding Authors:**

Rebecca Walwyn, Leeds Institute for Clinical Trials Research, University of Leeds, Leeds, United Kingdom, LS2 9JT.

Email: R.E.A.Walwyn@leeds.ac.uk

**Acknowledgements:** Rebecca Walwyn and Arpan Singh were funded by a National Institute for Health Research (NIHR) Advanced Fellowship (ref: NIHR301709). Rebecca Walwyn and Steven Gilmour were funded by an Engineering and Physical Sciences Research Council (EPSRC) Small Grant in Mathematical Sciences (ref: EP/W001020/1). Neil Corrigan was funded by an NIHR Doctoral Fellowship (ref: DRF-2018-11-ST2-079).

**Abstract**

It is recognised that treatment-related clustering should be allowed for in the sample size and analyses of individually-randomised parallel-group trials that evaluate therapist-delivered interventions such as psychotherapy. Here, interventions are a treatment factor, but therapists are not. If the aim of a trial is to separate effects of therapists from those of interventions, we propose that interventions and therapists should be regarded as two potentially interacting treatment factors (one fixed, one random) with a factorial structure. We consider the specific design where each therapist delivers each intervention (crossed therapist-intervention design), and the resulting therapist-intervention combinations are randomised to patients. We adopt a classical Design of Experiments (DoE) approach to propose a family of orthogonal factorial designs and their associated data analyses, which allow for therapist learning and centre too. We set out the associated data analyses using ANOVA and regression and report the results of a small simulation study conducted to explore the performance of the proposed randomisation methods in estimating the intervention effect and its standard error, the between-therapist variance and the between-therapist variance in the intervention effect. We conclude that more purposeful trial design has the potential to lead to better evidence on a range of complex interventions and outline areas for further methodological research.

## 1 Introduction

Kiesler [1] defined psychotherapy as lying "somewhere in the therapist and his behavio[u]r" (p. 128). Their behaviour is detailed in an intervention manual [2], which sets out the behaviours which are regarded as mandatory, optional, and prohibited [3]. The primary focus of trials is generally on estimating the effects of the intervention (e.g., the therapy or procedure). However, disentangling the effect of the therapists from the effect of the interventions has long been considered a challenge in individually-randomised, parallel-group trials [4 – 5]. Walwyn *et al* [6] have set out ten possible ways of designing a trial to separate the effects of the intervention and the therapist by including therapists as a factor when generating the randomisation sequence. These possibilities depend on the design of the complex intervention (the "therapist-intervention design" [6] here), which describes the structural relationships between the components of the complex intervention, irrespective of the randomisation. Five types of therapist-intervention design were proposed by Walwyn *et al* [6]. In the first, each therapist delivers all the interventions in a "crossed" design. Using this terminology, therapists are crossed with interventions, with every combination of each therapist and intervention being observed in the trial. Crossed designs have been suggested for psychotherapy trials where the aim is to control for enduring therapist traits [4], such as gender and personality, and where therapist allegiances to specific interventions are not a concern [7]. Crossed surgeon-intervention designs are the predominant design in surgical trials [8], as each surgeon usually undertakes all the procedures under investigation. Crossed therapist-intervention designs are commonly used in trials of psychotherapy, physiotherapy, occupational therapy and rehabilitation [6].

It has long been recognised that variation in average patient outcomes between therapists is expected [9]. More recently, it has been accepted that, likewise, patient outcomes are clustered by therapists, leading to intra-therapist correlation (for a review, see Walwyn and Roberts [10]). This consensus led to guidance on trial reporting [11]. There has been much discussion about whether [12 – 14] and when [15 – 18] therapists should be included as a fixed or random factor in analyses. In the early medical statistics literature, it was recommended that therapists are typically included as a random factor [10, 19 – 22] because there is often no interest in the outcomes of specific therapists in a trial. Instead, it is assumed that therapists are a random sample from a population of interest. Statistical models have been proposed (for a summary, see [10]). In all cases, the random effect of the therapist was allowed to vary by intervention, leading to use of the term "treatment-related clustering". For crossed designs, there was a consensus that a random coefficient model is appropriate, including a fixed effect for intervention, a random effect for therapist, an intervention-by-therapist random effect and a patient error [10, 22]. A method for calculating sample size was also proposed [10]. For psychotherapy, and a parallel literature on individually randomised group treatment (IRGT) trials, greater attention has been given to designs in which therapists are nested or partially nested in interventions (e.g., [10, 19, 20, 23 – 39]). This has left the literature on crossed therapist-intervention designs relatively underdeveloped.

In the statistical literature, it has largely been assumed that interventions are randomised to patients in an individually randomised, parallel-group design (see [21 – 22]). In this design, interventions are not randomised to therapists, and neither are therapists randomised to patients [10]. So, the intervention is regarded as a treatment factor, but the therapist is not. Here, we assume both the intervention and the therapist are treatment factors, with one combination of the levels of these two treatment factors (the "treatment") randomised to each patient (see Example A in [6]). We acknowledge that, in many trials, there are other contextual factors affecting the complex intervention delivered to patients that cannot be randomised. Therapists deliver interventions to patients sequentially in categorical or continuous time (see [40] for a review of the surgical literature on learning curves). While various measures of therapist learning are possible (e.g., number of previous patients seen, therapist caseload), the simplest is arguably time itself. Also, ongoing supervision and intervention policies can vary across centres. Centres may be recruiting centres, clinical services, or regions, but our focus is on the way in which therapists are grouped. Here, we suggest that time and centre are accounted for in the randomisation by including them as stratification (or "blocking") factors (see Figure 1 for a summary of the complex intervention designs considered in this paper).

[Insert Figure 1 about here]

There is a large body of statistical work on the Design of Experiments (DoE) and their data analysis that began with Fisher [41]. Perhaps apart from crossover designs, this has become separated from clinical trials, despite their common origins and its relevance. So, medical statisticians are largely unaware of developments in the DoE field and experts in the DoE are largely unaware of the problems encountered in clinical trials. Typically, in the DoE literature, treatment factors are all fixed. As such, adaptations are needed to include therapists as a random treatment factor. One general approach to DoE is that described in Bailey [42]. Instead of giving readers a menu of recognised designs, the approach gives them tools needed to develop experimental designs for the problem presented to them. Despite the flexibility of this approach, the first ten chapters of the book are limited to so-called "orthogonal" designs. These are, arguably, the theoretical ideal. From our perspective, their relative simplicity makes them a good place to start.

Using Bailey's [42] approach, the aim of this paper is to propose a family of increasingly complex, orthogonal factorial experimental designs, and their associated data analyses, for trials of therapist-delivered interventions, incorporating therapist learning and centre. We outline two motivating examples in Section 2. In Section 3 we introduce our (i) completely randomised, (ii) randomised block, and (iii) multicentre randomised block factorial designs, each illustrated with a theoretical example, inspired by the two motivating examples. In Section 4 we outline the resulting data analyses. Following Bailey [42], we give the analysis of variance (ANOVA) for each design first. Recognising that trialists are often more familiar with regression, the equivalent regression analysis is also given.

We summarise a small simulation study, and end with a discussion and some conclusions, in Sections 5 and 6, respectively.

## 2 Motivating Examples

Barkham *et al* [43 – 44] reported the design and main results of PRaCTICED, a trial embedded in the Sheffield Improving Access to Psychological Therapies (IAPT) service. The patients were recruited through general practices and there were four geographical districts in Sheffield, UK. For our purpose, however, this was a single centre trial, because the therapists were grouped within one IAPT service. In total, fifty therapists were involved. One of two interventions (Counselling for Depression (CfD), Cognitive Behavioural Therapy (CBT)) was randomised to 510 patients with block randomisation stratified by practice, using randomly varying blocks of size 2, 4 and 6. Therapists were non-randomly allocated to patients, and interventions started when a therapist became available in continuous time. Interventions were delivered by different samples of therapists (counsellors and therapists), so a nested therapist-intervention design was used, reflecting routine practice. It is conceivable, however, that counsellors could be trained to deliver both CfD and CBT, because some counsellors use an eclectic approach in practice. The advantages of a crossed relative to a nested design are twofold. First, it allows separation of the therapist and intervention-by-therapist random effects [6]. Second, fewer patients are required as the increase in sample size due to an intervention-by-therapist random effect is countered by a decrease in sample size required from the therapist random effect [10].

Shapiro *et al* [45 – 46] reported the design and results of the Second Sheffield Psychotherapy Project. This involved a single centre in the sense that the five therapists were all embedded in the research team. Patients self-referred or were referred by general practitioners or mental health services. By design, a combination of “duration” (8 or 16 weekly sessions), “method” (cognitive behavioural psychotherapy or psychodynamic-interpersonal psychotherapy) and “therapist” (1 to 5) was randomised to patients in each of three ranges of depression severity (low, moderate, high). So, in this trial, there was one blocking factor (severity) and three treatment factors (duration, method, therapist), even if allocation of therapists to patients was described as an extra randomisation (constrained by therapist availability, personal communication). The design was balanced; there were two replicates of each treatment combination, with each therapist treating two patients with each level of severity in each of the four combinations of duration and method. Therefore, this trial provides an example of an orthogonal factorial design in practice.

## 3 Proposed Factorial Designs

### *3.1 Completely randomised factorial design*

First consider a trial conducted in a single centre, where the interventions therapists deliver remain constant over time. Patients are the experimental units, and the treatments are all combinations of the levels of two treatment factors, $I$ (Interventions) and $T$ (Therapists), which have $n_I$ and $n_T$ levels, respectively. Assuming the $n_I n_T$ treatments are each replicated $n_R$ times, the total number of patients is $N = n_I n_T n_R$. We assume a continuous outcome is collected on patients at a single timepoint, so the observational units are also patients. Here, treatments are randomised to patients (see Figure 2a for a theoretical example).

[Insert Figure 2 about here]

Figure 2a is a two-tiered example of the randomisation diagrams introduced by Brien and Bailey [47] (see also [6]). The two boxes represent two sets of factors (or "tiers"), the set of treatments and the set of units. Following the convention of [47], the arrow representing the randomisation is from treatments to units. In practice, randomisation would be achieved by (i) numbering rows of a table 1,…, $N$, (ii) assigning each of the $n_I n_T$ treatments to $n_R$ rows in a "systematic design", (iii) choosing a random permutation of $\{1, \ldots, N\}$ by obtaining a sequence of distinct random numbers, (iv) assigning the random numbers to rows in order, (v) reordering the rows in the systematic design by reordering the random numbers in ascending order, and (vi) renumbering the rows in natural order, giving a random order of treatments for the rows (now patients) as they are recruited into the trial. This is equivalent to permuted block randomisation with a single block, equal in size to the total number of patients. It has been described as randomising the labels given to units [42]. Importantly, it leads to equal replication, which has useful properties for the analysis [42].

*3.2 Randomised block factorial design*

Next, we relax the assumption that the interventions therapists deliver remain constant over time. Interventions may commence when a therapist becomes available in continuous time or at the start of a "batch" in categorical time. Suppose patients are nested in blocks that are defined by time periods, creating the blocking factor $B$ (Batches), and assigned to $n_R$ replications of the combinations of levels of treatment factors $I$ and $T$ in each block. Factors $I$, $T$ and $B$ have $n_I$, $n_T$ and $n_B$ levels respectively, and the total number of patients is $N = n_I n_T n_B n_R$. As the precise treatment delivered to a patient can vary according to their position in their therapist's learning curve, one might consider randomising the time interventions commence, giving $n_I n_T n_B$ treatments. However, as in most trials, it is not possible to randomise Batches to Patients, there are $n_I n_T$ treatments (combinations of $I$ and $T$) here as before. Treatments are randomised to patients independently in each batch (see Figure 2b). Batches appears in the unit tier reflecting time being a blocking rather than a treatment factor in our design.

In Figure 2b, there are 64 patients in each batch, giving two replications of 32 treatments in each batch and 10 replications overall. Each therapist has four patients in each batch, two for each intervention, so over five batches, each therapist treats 20 patients. As such, $n_I = 2$, $n_T = 16$, $n_B = 5$ and $n_R = 2$, and the total number of patients is $n_I n_T n_B n_R = 320$. The randomisation would be done by (i) numbering rows in a table from 1,…, $n_I n_T n_R$ in each batch, (ii) assigning each of the $n_I n_T$ treatments to $n_R$ rows in each batch in a systematic design, (iii) choosing an independent random permutation for each batch of $\{1,\ldots,n_I n_T n_R\}$, and finally (iv) reordering and renumbering, as before, to give the random order of treatments for patients as they are recruited. This is equivalent to permuted block randomisation with $n_B$ blocks of fixed size $n_I n_T n_R$. Note that there are not $n_B n_R$ blocks of fixed size $n_I n_T$. This becomes important in the analysis (see [48]).

### *3.3 Multicentre randomised block factorial design*

Finally, we relax the assumption that the trial is conducted in a single centre. Here, patients are nested in the combinations of levels of two blocking factors, Centres ($C$) and Batches ($B$). Factors $C$ and $B$ have $n_C$ and $n_B$ levels, respectively, creating $n_C n_B$ blocks. Often therapists will already have a centre, so in this design $n_T$ Therapists are nested *in each* of $n_C$ Centres; it is not possible to assign Therapists to Centres. Previously $n_T$ denoted the total number of therapists. Patients are assigned to $n_R$ replications of the combinations of the *relevant* levels of treatment factors $I$ and $T$ in each batch in each centre, so the total number of patients is $N = n_I n_T n_C n_B n_R$. As the precise treatment delivered to each patient also varies according to the centre a therapist belongs to, one might consider randomising a combination of centres and batches, giving $n_I n_C n_B n_T$ treatments. However, usually there will only be $n_I n_C n_T$ treatments, as there are $n_C n_T$ therapists, and each intervention occurs with every therapist. So, the number of centres contributes to the number of treatments. In each combination of Centres and Batches, treatments (combinations of $I$ and $T$ nested in $C$) are independently randomised to patients (see Figure 2c). As Therapists and Patients are nested in Centres, the double lines connecting Centres across tiers indicates that Centres is a factor that spans both tiers. Although this situation is common in clinical trials (see [6]), it was not faced in any of the relevant Brien-Bailey papers [49 – 55].

In Figure 2c, there are 16 treatments and 160 patients in each of six centres, so 96 treatments and 960 patients overall. There are 32 patients and two replications of the 16 treatments in each of 30 centre-batch combinations, and two replications of the 96 treatments in each batch. Each of 48 therapists has four patients in each batch, two for each intervention. So, in total each therapist treats 20 patients. Hence, $n_I = 2$, $n_C = 6$, $n_B = 5$, $n_T = 8$ and $n_R = 2$, so $N = 960$. The randomisation would be done by (i) numbering rows in a table from $1,\ldots, n_I n_T n_R$ in each combination of Centres and Batches, assigning $n_I n_T$ treatments to each of $n_R$ patients in a systematic design, (ii) choosing an independent random permutation of $\{1,\ldots n_I n_T n_R\}$ for each centre-batch combination and (iii) reordering and renumbering as before to give the random order of treatments for patients as they are recruited. This is like stratified permuted block randomisation with $n_C n_B$ blocks of size $n_I n_T n_R$, except that there is no overlap in the therapists from one centre to the next. There are not $n_C n_B n_R$ blocks of fixed size $n_I n_T$.

## 4 Proposed Data Analyses

### *4.1 Hasse Diagrams*

Figure 2 gives a concise way of describing the randomisation. However, to understand the design sufficiently to determine the data analysis, it is informative to show the unit and treatment factors in Hasse diagrams (see [42] for more on factors, finer and coarser relationships between factors, infima and suprema, notation, and Hasse diagrams). Here, all the relationships between all the factors in the design are made clear. The technical detail in Sections 4.1 and 4.2 can be skipped on first reading this paper. It is included to highlight how we have adapted the approach in [42]. Specifically, we recommend that, where the treatment factors contribute fixed and random effects, it is helpful to give separate Hasse diagrams for fixed and random factors, rather than for unit and treatment structures. Figure 3 gives the Hasse diagrams for our proposed designs. To clearly distinguish these Hasse diagrams from those in [42], we recommend using white diamonds to represent fixed factors and black diamonds for random factors, whether they are part of the unit or treatment structure. Note Bailey [42] defines two special factors, the Universal factor $U$ and the Equality factor $E$. The Universal factor is constant, making no distinctions between the units. It has a single level, for all of the units, and serves as the basis for estimating the overall mean. The Equality factor is different for each unit, so it has as many levels as there are units.

[Insert Figure 3 about here]

Note also that in this paper, we will make frequent use of the concept of the infimum of two factors. Given two factors $F$ and $G$ on the same set, their 'infimum' is the factor denoted $F \wedge G$ each of whose levels is a combination of a level of $F$ with a level of $G$. The supremum $F \vee G$ of factors $F$ and $G$ is more complicated to define. If two units $\alpha$ and $\beta$ have the same level of $F$ or the same level of $G$ then they must have the same level of $F \vee G$. The supremum $F \vee G$ is the unique factor with the smallest number of levels that satisfies this criterion. It is apparent from Figure 3a that the random factors are all nested and that the fixed factors are unstructured in our completely randomised factorial design.

In our randomised block factorial design, in addition to the infimum $I \wedge T$ of factors $I$ and $T$, we consider infima of the treatment and blocking factors, $I \wedge B$, $T \wedge B$ and $I \wedge T \wedge B$, because Batches constitute part of the complex intervention design, making interactions between treatment and blocking factors plausible. Factors $B$, $I \wedge B$, $T \wedge B$ and $I \wedge T \wedge B$ all contribute random effects. Next, we consider suprema of all pairs of random factors because the supremum of any two random factors must be included in the list of random factors used to identify the eigenspaces of the variance-covariance matrix of the patient outcomes, $Cov(\boldsymbol{Y})$. Unusually, this shows that the supremum of the random factors $I \wedge T$ and $I \wedge B$ is aliased with the fixed factor $I$. Yet, factor $I$ does not contribute a new variance parameter. We call its contribution a "dependent random effect". As a Hasse diagram of the random structure should include all suprema of the random factors, we include factor $I$ in Figure 3b(i) but give it a special symbol to distinguish it both from more traditional random factors and from the factor $I$ considered a fixed factor.

In our multicentre randomised block factorial design, in addition to $I \wedge T$, $I \wedge B$, $T \wedge B$ and $I \wedge T \wedge B$ we consider the infima $I \wedge C$, $C \wedge B$ and $I \wedge C \wedge B$, since Centre also constitutes part of the complex intervention design, making interactions plausible. Here, factors $C$, $I \wedge C$, $C \wedge B$ and $I \wedge C \wedge B$ all contribute random effects. Next, we consider suprema of all pairs of random factors and find that the supremum of the random factors $I \wedge C$ and $I \wedge B$ and the random factors $I \wedge B$ and $I \wedge T$ is aliased with the fixed factor $I$. This explains why we have included factor $I$ in Figure 3c(i).

*4.2 Analysis of Variance (ANOVA)*

The standard linear model for the outcome variable $Y_\omega$ on patient $\omega$ is a sum of several terms, one for each relevant factor. A factor with fixed effects contributes a (unknown in advance) constant for each of its levels. A factor with random effects contributes a random variable for each of its levels. All the random variables have zero mean and are mutually independent, and all the random variables defined by levels of the same factor have the same variance. For each design in Figure 2, the corresponding Hasse diagram in Figure 3 was used to construct the first three columns in the ANOVA table. These are shown in Table 1.

[Insert Table 1 about here]

*4.2.1* Strata

Figures 3a(i), 3b(i) and 3c(i) were used to obtain the eigenspaces of $Cov(\boldsymbol{Y})$. These are the "strata" in Table 1 (the first column). Each stratum is associated with a $W$ subspace of the $N$-dimensional vector space $V$ associated with the set of patients, with degrees of freedom given by its dimension. There is a stratum for each factor in Figures 3a(i), 3b(i) and 3c(i). Importantly, the $W$ subspaces consist of all vectors associated with each factor that are orthogonal to vectors associated with other factors ([42] gives more details on vector spaces and the null ANOVA). So $V$ can be expressed as the orthogonal direct sum of the $W$ subspaces associated with the design.

Our completely randomised factorial design is relatively straightforward, since it has an orthogonal block structure in the sense defined by Houtman and Speed [56]. In contrast, the block structures in our randomised block and multicentre randomised block factorial designs are not orthogonal under this [56] definition, because factor $I$ contributes a dependent random effect in each case. Specifically, for the randomised block factorial design, the intersection between the vector spaces associated with random factors $I \wedge T$ and $I \wedge B$ is the vector space associated with factor $I$. Also, these vector spaces are contained in the vector space associated with random factor $I \wedge T \wedge B$. As such, the vector spaces associated with $I \wedge T$, $I \wedge B$ and $I \wedge T \wedge B$ are not orthogonal to one another. Similarly, for the multi-centre randomised block factorial design, the intersection between the vector spaces associated with random factors $I \wedge C$ and $I \wedge B$ is the vector space associated with factor $I$. Also, the vector space associated with $I \wedge C$ is contained in that associated with $I \wedge T$, the vector space associated with $I \wedge B$ is contained in that associated with $I \wedge C \wedge B$, and the vector spaces associated with $I \wedge C$ and $I \wedge B$ are both contained in that associated with $I \wedge T \wedge B$. As such, here, the vector spaces associated with factors $I \wedge C$, $I \wedge B$, $I \wedge T$, $I \wedge C \wedge B$ and $I \wedge T \wedge B$ are not orthogonal to one another (more detail on this can be found in Supplementary File A).

#### *4.2.2* Sources of variation

Figures 3a(ii), 3b(ii) and 3c(ii) were used to give subspaces of the strata. These are the "sources of variation" in Table 1 (the second column). Here, vectors associated with factor $I$ are orthogonal to those associated with the mean. The appropriate stratum for fixed factor $I$ is identified by finding the supremum of all random factors which occur below factor $I$ in a combined Hasse diagram (see Supplementary File B). In the completely randomised factorial design, the appropriate stratum is that associated with factor $I \wedge T$. This stratum is subdivided into a subspace associated with the fixed factor $I$ and a residual, the latter sometimes being referred to as the intervention-by-therapist interaction.

In the randomised block factorial design, the relevant stratum for fixed factor $I$ is now the $W$ subspace associated with factor $I$, $W_I$. However, the fixed effect of $I$ takes up the whole of $W_I$, so no part of $W_I$ can give a residual mean square whose value can be used to provide an unbiased estimator of the appropriate residual. Instead, a linear combination of random factors is indicated. This is the case for the multicentre randomised block factorial design as well, except that the linear combination of random factors indicated differs.

#### *4.2.3* Degrees of Freedom

The degrees of freedom are given in the third column in Table 1. The degrees of freedom associated with the strata are given on the right-hand side, and those relating to the fixed structure and residuals (where applicable) are given on the left-hand side. Any residual degrees of freedom are calculated by subtraction from the relevant stratum total.

#### *4.2.4* Expected Mean Squares

The final column of Table 1 gives the expected mean squares. As $I$ is the only non-trivial factor with fixed effects, we assume there are unknown real numbers $\tau_i$ for each level $i$ of $I$ so the expectation of the random outcome variable $Y_\omega$ on patient $\omega$ is $\tau_i$ if factor $I$ takes level $i$ on patient $\omega$. If all random variables $Y_\omega$ are put into the random vector $\boldsymbol{Y}$ of length $N$, and all the values $\tau_{I(\omega)}$ are put into a real vector $\boldsymbol{\tau}$ of length $N$, then $\mathbb{E}(\boldsymbol{Y}) = \boldsymbol{\tau}$. As the Universal factor $U$ contributes a fixed effect, which is the overall mean $\bar{\tau}$ of all the entries of $\boldsymbol{\tau}$, we let $\boldsymbol{\tau}_0$ be the vector of length $N$ with all entries equal to $\bar{\tau}$ and put $\boldsymbol{\tau}_I = \boldsymbol{\tau} - \boldsymbol{\tau}_0$. Then $\mathbb{E}(\boldsymbol{Y}) = \boldsymbol{\tau}_0 + \boldsymbol{\tau}_I$ and $\boldsymbol{\tau}_0 \in W_0$ and $\boldsymbol{\tau}_I \in W_I$. Note that $\|\boldsymbol{\tau}_0\|^2$ and $\|\boldsymbol{\tau}_I\|^2$ are the sums of squares of $\boldsymbol{\tau}_0$ and $\boldsymbol{\tau}_I$, respectively. The $\xi$ are the eigenvalues of $Cov(\boldsymbol{Y})$ (see Supplementary File C for details). These make it clear which residual is appropriate for testing the fixed effect of $I$ and what the appropriate F tests of the random effects are (for the latter, see Supplementary File D).

In a completely-randomised factorial design, the appropriate residual for testing the fixed effect of $I$ is the one whose expected mean square is $\xi_{I \wedge T}$. Importantly, it is not the residual whose expected mean square is $\xi_E$ or $(15\xi_{I \wedge T} + 288\xi_E)/303$, both of which are smaller than $\xi_{I \wedge T}$. If the therapist variance $\sigma_T^2 = 0$ then $\xi_T = \xi_{I \wedge T}$ and so $MS_T / RMS_{I \wedge T}$ has an F-distribution on 15 and 15 degrees of freedom in the running example, where $MS_T$ is the observed mean square for $T$ and $RMS_{I \wedge T}$ the observed residual mean square for $I \wedge T$. If the therapist variance in the intervention effect $\sigma_{I \wedge T}^2 = 0$, then $\xi_{I \wedge T} = \xi_E$ and so $RMS_{I \wedge T} / MS_E$ has an F-distribution on 15 and 288 degrees of freedom in the running example, where $MS_E$ is the observed mean square for $E$.

In a randomised block factorial design, the appropriate residual for testing the fixed effect of $I$ is one whose expected mean square is a linear combination of eigenvalues of $Cov(\boldsymbol{Y})$, $\xi_{I \wedge T} + \xi_{I \wedge B} - \xi_{I \wedge T \wedge B}$. As such, the eigenvalues of $Cov(\boldsymbol{Y})$ are linearly dependent. That is, the eigenvalue for $W_I$ is now a linear combination of those for subspaces $W_{I \wedge T}$, $W_{I \wedge B}$ and $W_{I \wedge T \wedge B}$. Though the block structure is not orthogonal, the linear dependence on three other eigenvalues means that a linear combination of the mean squares for $W_{I \wedge T}$, $W_{I \wedge B}$ and $W_{I \wedge T \wedge B}$ from the ANOVA can be used to provide an unbiased estimator of the appropriate residual mean square. Satterthwaite [57] proposed that if a sample variance is a linear combination of independent mean squares, its exact distribution can be approximated by a chi-square distribution whose effective degrees of freedom are given by a specified formula involving the relevant mean squares and their degrees of freedom. The degrees of freedom for the linear combination of mean squares estimating the appropriate residual $\xi_{I \wedge T} + \xi_{I \wedge B} - \xi_{I \wedge T \wedge B}$ can be approximated by

$$df_I = \frac{[MS_{I \wedge T} + MS_{I \wedge B} - MS_{I \wedge T \wedge B}]^2}{\frac{[MS_{I \wedge T}]^2}{(n_I - 1)(n_T - 1)} + \frac{[MS_{I \wedge B}]^2}{(n_I - 1)(n_B - 1)} + \frac{[-MS_{I \wedge T \wedge B}]^2}{(n_I - 1)(n_T - 1)(n_B - 1)}} \qquad (1)$$

where $MS_{I \wedge T}$, $MS_{I \wedge B}$ and $MS_{I \wedge T \wedge B}$ are the observed mean squares for factors $I \wedge T$, $I \wedge B$ and $I \wedge T \wedge B$, respectively. This result can be used to provide an approximate F-test for the effect of $I$.

A similar situation arises in a multicentre randomised block factorial design, where the appropriate residual for testing the fixed effect of $I$ is one whose expected mean square is $\xi_{I \wedge C} + \xi_{I \wedge B} - \xi_{I \wedge C \wedge B}$. The eigenvalues of $Cov(\boldsymbol{Y})$ are again linearly dependent, with the eigenvalue for $W_I$ a linear combination of those for subspaces $W_{I \wedge C}$, $W_{I \wedge B}$ and $W_{I \wedge C \wedge B}$. The degrees of freedom for the linear combination of mean squares estimating the appropriate residual $\xi_{I \wedge C} + \xi_{I \wedge B} - \xi_{I \wedge C \wedge B}$ can be approximated by

$$df_I = \frac{[MS_{I \wedge C} + MS_{I \wedge B} - MS_{I \wedge C \wedge B}]^2}{\frac{[MS_{I \wedge C}]^2}{(n_I - 1)(n_C - 1)} + \frac{[MS_{I \wedge B}]^2}{(n_I - 1)(n_B - 1)} + \frac{[-MS_{I \wedge C \wedge B}]^2}{(n_I - 1)(n_C - 1)(n_B - 1)}} \qquad (2)$$

where $MS_{I\wedge C}$, $MS_{I\wedge B}$ and $MS_{I\wedge C\wedge B}$ are the observed mean squares for factors $I \wedge C$, $I \wedge B$ and $I \wedge C \wedge B$. This can be used to provide an approximate F-test for the fixed effect of $I$.

*4.3 Regression*

In clinical trials, the linear model is often written as a regression model (see for example [19 – 22]).

*4.3.1* Completely-Randomised Factorial Design

Here, the regression model for a continuous outcome for patient $i$ $(i = 1, \dots, 320)$ nested in therapist $j$ $(j = 1, \dots, 16)$ can be given as the following random coefficient model,

$$Y_{ij} = \delta_0 + \delta_1 I_{ij} + u_{1j} + v_{1j} I_{ij} + e_{ij} \qquad (3)$$

where the notation $Y_{ij}$ means that patients and therapists are numbered from 1 to 320 and 1 to 16, respectively. In this model, $I_{ij}$ are indicator variables for the intervention assigned to patient $i$ and therapist $j$. The $I_{ij}$ are not dummy (0, 1) variables. They are treatment contrasts with coefficients that sum to zero. For two equally-replicated interventions, a single dummy variable may be replaced by an effect-coded (–1, 1) one; –1 for the control and 1 for the experimental intervention. Here, $\delta_0$ is the overall mean outcome, $\delta_1$ is the difference between the mean of all outcomes for each intervention and $\delta_0$, $u_{1j}$ is a random effect around $\delta_0$ for the $j$th therapist, $v_{1j}$ is a random effect around $\delta_1$ for the $j$th therapist, and $e_{ij}$ is the patient level error. Here, $u_{1j} \sim N(0, \sigma_{u1}^2)$, $v_{1j} \sim N(0, \sigma_{v1}^2)$ and $e_{ij} \sim N(0, \sigma_e^2)$ and $u_{1j}, v_{1j}$ and $e_{ij}$ are all mutually independent, so $\sigma_{u1}^2$ is the between-therapist variance, $\sigma_{v1}^2$ is the between-therapist variance in the intervention effect and $\sigma_e^2$ is the patient-level variance. Note that in a linear mixed model normality of random effects is nearly always assumed; this is not so in ANOVA.

To generalise Model (3) to $n_I$ interventions, the contrast $I_{ij}$ is replaced by $n_I - 1$ orthogonal contrasts, $I_{1ij}, \dots, I_{(n_I-1)ij}$. Each of these contrasts has coefficients that sum to zero across all patients, and each pair of contrasts is such that the product of their coefficients sums to zero across all patients. For three interventions, the two contrasts may be one coded 1/2, 1/2 and –1 and one coded 1, –1 and 0 for interventions 1, 2 and 3. Here, $\delta_0$ remains the overall mean outcome, $\delta_{11}$ is the difference between $\delta_0$ and the mean of all outcomes for intervention 3, and $\delta_{21}$ is half the difference between the mean of all outcomes for interventions 1 and 2. The variance parameters are unchanged. Except multiplying the coefficients of either contrast by a constant, and/or relabelling the interventions, there is no other meaningful orthogonal option for three interventions. For more interventions, the set of orthogonal contrasts should match any plausible structure in the interventions.

Regardless of the number of interventions, it is possible to write the eigenvalues $\xi_E$, $\xi_{I\wedge T}$, $\xi_T$ and $\xi_0$ in the ANOVA in terms of the variance components $\sigma_e^2$, $\sigma_{u1}^2$ and $\sigma_{v1}^2$ from the regression as follows,

$$\xi_E = \sigma_e^2 \qquad (4)$$

$$\xi_{I\wedge T} = \sigma_{v1}^2 n_R + \sigma_e^2 \qquad (5)$$
$$\xi_T \ = n_I n_R \sigma_{u1}^2 + n_R \sigma_{v1}^2 + \sigma_e^2 \qquad (6)$$
$$\xi_0 \ = 0 + n_I n_R \sigma_{u1}^2 + n_R \sigma_{v1}^2 + \sigma_e^2 \qquad (7).$$

In this variance components model, there is a sum of random variables, each contributing a different variance, all of which are non-negative, and which together give the variance-covariance matrix of the outcomes. This model can be generalised by defining the variance-covariance matrix directly, so that no variance is negative, but the correlation between outcomes can be. If so-called "negative variance components" are allowed, the regression defined in (3) and the ANOVA (Table 1) are identical.

*4.3.2* Randomised Block Factorial Design

Here, with $n_I = 2$, the regression model for a continuous outcome $Y_{ijk}$ for patient $i$ $(i = 1, \dots, 320)$ nested in a combination of therapist $j$ $(j = 1, \dots, 16)$ and batch $k$ $(k = 1, \dots 5)$ can be written as a cross-classified or crossed random effects model as follows,

$$Y_{ijk} = \delta_0 + \delta_1 I_{ijk} + u_{1j} + u_{2k} + u_{3jk} + v_{1j} I_{ijk} + v_{2k} I_{ijk} + v_{3jk} I_{ijk} + e_{ijk} \qquad (8)$$

where the notation $Y_{ijk}$ means that patients, therapists and batches are numbered from 1 to 320, 1 to 16, and 1 to 5, respectively. Here, $I_{ijk}$ is an indicator for the intervention assigned to patient $i$ nested in therapist $j$ and batch $k$. As before, the $I_{ijk}$ represent an orthogonal treatment contrast, coded 1 and –1 for levels 1 and 2 of factor $I$, so $\delta_0$ is the overall mean, $\delta_1$ is the mean intervention effect (defined as the difference between the mean of all outcomes for each intervention and $\delta_0$), $u_{1j}$ is a random effect around $\delta_0$ for the $j$th therapist, $u_{2k}$ is a random effect around $\delta_0$ for the $k$th batch, and $u_{3j}$ is a random effect around $\delta_0 + u_{2k}$ for the $j$th therapist. Also, $v_{1j}$ is a random effect around $\delta_1$ for the $j$th therapist, $v_{2k}$ is a random effect around $\delta_1$ for the $k$th batch, $v_{3jk}$ is a random effect around $\delta_1 + v_{2k}$ for the $j$th therapist, and $e_{ijk}$ is the patient level error. Here, $u_{1j} \sim N(0, \sigma_{u1}^2), u_{2k} \sim N(0, \sigma_{u2}^2)$, $u_{3j} \ \sim N(0, \sigma_{u3}^2)$, $v_{1j} \sim N(0, \sigma_{v1}^2)$, $v_{2k} \sim N(0, \sigma_{v2}^2)$, $v_{3j} \ \sim N(0, \sigma_{v3}^2)$ and $e_{ijk} \sim N(0, \sigma_e^2)$ where $u_{1j}$, $u_{2k}$, $u_{3jk}$, $v_{1j}$, $v_{2k}$ $v_{3jk}$ and $e_{ijk}$ are all mutually independent. Thus, $\sigma_{u1}^2$ is the between-therapist variance, $\sigma_{u2}^2$ is the between-batch variance, $\sigma_{u3}^2$ is the between-therapist-by-batch interaction, $\sigma_{v1}^2$ is the between-therapist variance in the intervention effect, $\sigma_{v2}^2$ is the between-batch variance in the intervention effect, $\sigma_{v3}^2$ is the therapist-by-batch-by-intervention interaction and $\sigma_e^2$ is the patient-level variance. Model (8) can be generalised to $n_I$ interventions in the same way as Model (3).

*4.3.3* Multicentre Randomised Block Factorial Design

Here, with $n_I = 2$, the regression model for continuous outcome $Y_{ijkl}$ for patient $i$ $(i = 1, \dots, 960)$ nested in a combination of therapist $j$ $(j = 1, \dots, 8)$ and batch $k$ $(k = 1, \dots 5)$, where $j$ is nested in centre $l$ $(l = 1, \dots 6)$ and $k$ is crossed with centre $l$, can be written as a cross-classified model as,

$$Y_{ijkl} = \delta_0 + \delta_1 I_{ijkl} + u_{1j} + u_{2k} + u_{3jk} + u_{4l} + u_{5kl} + v_{1j}I_{ijkl} + v_{2k}I_{ijkl} + v_{3jk}I_{ijkl} + v_{4l}I_{ijkl} + v_{5kl}I_{ijkl} + e_{ijkl} \quad (9)$$

where the notation $Y_{ijkl}$ means that patients, therapists, batches and centres are numbered from 1 to 960, 1 to 48, 1 to 5, and 1 to 6, respectively. Here, the $I_{ijkl}$ are indicators for the intervention assigned to patient $i$ nested in therapist $j$, batch $k$ and centre $l$. The $I_{ijkl}$ again represent a treatment contrast, coded 1 and $-1$ for levels 1 and 2 of factor $I$. Hence, $\delta_0$ is the overall mean outcome, $\delta_1$ is the mean intervention effect (defined as the difference between the mean of all outcomes for each intervention and $\delta_0$), $u_{1j}$ is a random effect around $\delta_0$ for the $j$th therapist, $u_{2k}$ is a random effect around $\delta_0$ for the $k$th batch, $u_{3jk}$ is a random effect around $\delta_0 + u_{2k}$ for the $j$th therapist, $u_{4l}$ is a random effect around $\delta_0$ for the $l$th centre and $u_{5kl}$ is a random effect around $\delta_0 + u_{2k}$ for the $l$th centre. Additionally, $v_{1j}$ is a random effect around $\delta_1$ for the $j$th therapist, $v_{2k}$ is a random effect around $\delta_1$ for the $k$th batch, $v_{3jk}$ is a random effect around $\delta_1 + v_{2k}$ for the $j$th therapist, $v_{4l}$ is a random effect around $\delta_1$ for the $l$th centre, $v_{5kl}$ is a random effect around $\delta_1 + v_{2k}$ for the $l$th centre and $e_{ijkl}$ is the patient level error. Thus, $u_{1j} \sim N(0, \sigma_{u1}^2)$, $u_{2k} \sim N(0, \sigma_{u2}^2)$, $u_{3j} \sim N(0, \sigma_{u3}^2)$, $u_{4l} \sim N(0, \sigma_{u4}^2)$, $u_{5kl} \sim N\left(0, \sigma_{u5}^2\right)$, $v_{1j} \sim N(0, \sigma_{v1}^2)$, $v_{2k} \sim N(0, \sigma_{v2}^2)$, $v_{3j} \sim N(0, \sigma_{v3}^2)$, $v_{4l} \sim N(0, \sigma_{v4}^2)$, $v_{5kl} \sim N\left(0, \sigma_{v5}^2\right)$ and $e_{ijkl} \sim N(0, \sigma_e^2)$ where $u_{1j}$, $u_{2k}$, $u_{3jk}$, $u_{4l}$, $u_{5kl}$, $v_{1j}$, $v_{2k}$, $v_{3jk}$, $v_{4l}$, $v_{5kl}$ and $e_{ijkl}$ are all mutually independent. Hence, $\sigma_{u1}^2$ is the between-therapist variance, $\sigma_{u2}^2$ is the between-batch variance, $\sigma_{u3}^2$ is the between-therapist-by-batch interaction, $\sigma_{u4}^2$ is the between-centre variance, $\sigma_{u5}^2$ is the between-centre-by-batch interaction, $\sigma_{v1}^2$ is the between-therapist variance in the intervention effect, $\sigma_{v2}^2$ is the between-batch variance in the intervention effect, $\sigma_{v3}^2$ is the therapist-by-batch-by-intervention interaction, $\sigma_{v4}^2$ is the between-centre variance in the intervention effect, $\sigma_{v5}^2$ is the centre-by-batch-by-intervention interaction, and $\sigma_e^2$ is the patient-level variance. Model (9) can be generalised to $n_I$ interventions in the same way as Model (3).

The relationships between the eigenvalues from the ANOVA and the variance components from the regression are given in Supplementary File E for the randomised block and multicentre randomised block factorial designs.

### *4.4 Analysis in R*

The analysis can be done either via the `aov` command or the `lmer` command in R (Example output is given in Supplementary File F where factor $I$ has two or three levels). The following code can be used to obtain the ANOVA for the completely randomised factorial design:

```
aov1 <- aov(y~I+Error(T+I:T), contrasts=list(I=contr.sum), data=CrossedDataset1)
summary(aov1)
coefficients(aov1)
print(model.tables(aov1,"means"))
```

These commands give the ANOVA table, an F test for the fixed effect of $I$ and estimates of $\xi_T$, $\xi_{I\wedge T}$

and $\xi_E$. They give estimates of $\delta_0$ and $\delta_1$ from Model (3). To get F tests for the random effects of factors $T$ and $I \wedge T$, it is necessary to calculate the F statistics by hand using the estimates of $\xi_T$, $\xi_{I\wedge T}$ and $\xi_E$ to get the p-value.

The following code gives the regression using residual maximum likelihood (REML):

```
reg1 <- lmer(y~I+(1|T)+(1|I:T), contrasts=list(I=contr.sum), data=CrossedDataset1)
summary(reg1)
ls_means(reg1, which = "I", pairwise = TRUE)
ranova(reg1)
```

The regression table, a t-test for the fixed effect of $I$ and estimates of $\sigma^2_{u1}$, $\sigma^2_{v1}$ and $\sigma^2_e$ can be obtained, along with the least squares mean difference for the fixed effect of $I$ and likelihood ratio tests for the random effects $\sigma^2_{u1}$ and $\sigma^2_{v1}$. This analysis constrains the correlation between $\sigma^2_{u1}$ and $\sigma^2_{v1}$ to be zero. The contrast specified also ensures the correlation between estimates of $\delta_0$ and $\delta_1$ is zero.

The ANOVA for the randomised block factorial design can be obtained using the following code:

```
aov2 <- aov(y~I+Error(I*T*B), contrasts=list(I=contr.sum), data=CrossedDataset2)
```

The fixed effect of $I$ appears in the fixed and random parts of the model. If it is omitted from the random part, R allocates the fixed effect of $I$ to either the $W_{I\wedge T}$ or $W_{I\wedge B}$ stratum, depending on whether I:T or I:B is included in the random part of the model first. An F test is given with the observed residual mean square relating to $\xi_{I\wedge T}$ or $\xi_{I\wedge B}$ not $\xi_{I\wedge T} + \xi_{I\wedge B} - \xi_{I\wedge T\wedge B}$. The observed residual variance for $I$ can be calculated by hand using the estimates for $\xi_{I\wedge T}, \xi_{I\wedge B}$ and $\xi_{I\wedge T\wedge B}$. The approximate F statistic is $MS_I/(MS_{I\wedge T} + MS_{I\wedge B} - MS_{I\wedge T\wedge B})$. Numerator degrees of freedom can be taken from the ANOVA table and denominator degrees of freedom from Equation (1). This gives an approximate F test for the main effect of $I$. Approximate F tests for the random effects of $T$ and $B$ can be found using Supplementary File D. Exact F tests for the random effects of $I \wedge T, I \wedge B,\ T \wedge B$ and $I \wedge T \wedge B$ are straightforward, but also need to be done by hand.

The following code gives the regression for the randomised block factorial design:

```
reg2 <- lmer(y~I+(1|T)+(1|B)+(1|T:B)+(1|I:T)+(1|I:B)+(1|I:T:B), contrasts=list(I=contr.sum),
         data=CrossedDataset2)
```

Here, the approximate t-test for the fixed effect of $I$ using a Satterthwaite approximation for the degrees of freedom is given directly.

The ANOVA for a multicentre randomised block factorial design can be obtained using:

```
aov3 <- aov(y~I+Error(B+I+C+T+I:B+C:B+T:B+I:C+I:T+I:C:B+I:T:B), contrasts=list(I=contr.sum),
        data=CrossedDataset3)
```

The effect of $I$ appears in the fixed and random parts of the model. If it is omitted from the random part, R allocates the fixed effect of $I$ to either the $W_{I\wedge T}$, $W_{I\wedge C}$ or $W_{I\wedge B}$ stratum, depending on whether I:T, I:C or I:B is included in the random part of the model first. An F test is thus given with the observed residual mean square relating to $\xi_{I\wedge T}$, $\xi_{I\wedge B}$ or $\xi_{I\wedge C}$ not $\xi_{I\wedge C} + \xi_{I\wedge B} - \xi_{I\wedge C\wedge B}$. The appropriate residual variance for $I$ can be calculated by hand using estimates for $\xi_{I\wedge C}$, $\xi_{I\wedge B}$ and $\xi_{I\wedge C\wedge B}$. The approximate F statistic is $MS_I/(MS_{I\wedge C} + MS_{I\wedge B} - MS_{I\wedge C\wedge B})$. Numerator degrees of freedom can be taken from the ANOVA table, denominator degrees of freedom come from Equation (2). This gives an approximate F test for the main effect of $I$. Approximate F tests for the random effects of $C, B$, $I \wedge C, T$, and $C \wedge B$ can be found using Supplementary File D. Exact F tests for the random effects of $I \wedge B, I \wedge T, I \wedge C \wedge B, T \wedge B$ and $I \wedge T \wedge B$ are straightforward, but again need to be done by hand.

The following code gives the regression for the multicentre randomised block factorial design:

```
reg3 <- lmer(y~I+(1|T)+(1|B)+(1|T:B)+(1|C)+(1|C:B)+(1|I:T)+(1|I:B)+(1|I:T:B)+(1|I:C)+
          (1|I:C:B), contrasts=list(I=contr.sum), data=CrossedDataset3)
```

Here, an approximate t-test for the fixed effect of $I$ using a Satterthwaite approximation for the degrees of freedom is again given directly.

## 5 Simulation Study

A simulation study was conducted to explore the performance of the proposed randomisation methods on the estimation of the intervention effect $\delta_1$ and its standard error, the between-therapist variance $\sigma_{u1}^2$, and the between-therapist variance in the intervention effect $\sigma_{v1}^2$, from Models (3), (8) and (9) (see https://github.com/arpan52/Randomising-IxT-combinations-to-patients). Specifically, we compared the following five methods: (i) non-randomly assign therapists to patients *after* randomising interventions to patients; (ii) non-randomly assign therapists to patients *before* randomising interventions to patients; (iii) blocking the randomisation of interventions to patients by therapists; (iv) randomising the combination of therapist and intervention to patients (our proposed method); and (v) randomising interventions to patients, then randomising therapists to patients, blocking the latter by the randomised intervention. We made four comparisons: 1) method (i) versus (ii); 2) method (ii) versus (iii); 3) method (iii) versus (iv); and 4) method (iv) versus (v). For each comparison, 10000 simulated datasets were generated for each of the three illustrative examples taken from Figure 2. We generated the systematic design then fixed the true values of the fixed and random effects for each example as summarised in Table 2 below.

[Insert Table 2 about here]

We varied the degree to which the non-random assignment of therapists to patients was determined by matching patients to therapists based on a baseline characteristic of the patient that was predictive of patient outcome (for each therapist, the average value of this baseline covariate was assumed to follow a Normal distribution with mean $\delta_2$ and variance one (i.e., $N(\delta_2, 1)$). For all five methods, in the systematic design (see Sections 3.1, 3.2 and 3.3) $\delta_2$ was set to be either zero or 0.2. When $\delta_2 = 0$, the baseline covariate was excluded from the data generation process. These values were chosen because our preliminary simulations indicated that once $\delta_2$ is non-zero, the magnitude of its expected value has minimal impact on the estimation of $\delta_1$ or the associated inference. We further varied the degree to which the non-random assignment of therapists to patients was influenced by knowledge of the intervention assignment, again using a baseline characteristic of the patient that was predictive of patient outcome (for each therapist, the average value of this baseline covariate was assumed to follow a Normal distribution with mean $\delta_3$ and variance one (i.e., $N(\delta_3, 1)$). If therapists were assigned to patients before (or at the same time as) interventions were randomised to patients (i.e., methods (ii), (iii) and (iv)), $\delta_3$ was set to be zero (the baseline covariate was excluded from the data generation process), assuming that allocation of interventions to patients was perfectly concealed at the point where therapists were assigned to patients. Where therapists were assigned to patients after interventions were randomised to patients (i.e., methods (i) and (v)), $\delta_3$ was set to be either zero or 0.2 in the systematic design for similar reasons as for $\delta_2$.

Then we generated the randomised design. First, the randomisation method for allocating therapists and interventions to patients in batches and centres was determined for each method $i \in \{1, \dots, 5\}$ and example $j \in \{1, 2, 3\}$, according to design $D_{ij}$. Second, continuous patient outcomes were generated consistent with the example (i.e., with (3), (8) and (9) for example $j = 1, 2$ and 3 respectively), using the specified model parameters (see Table 2) in addition to $\delta_2$ and $\delta_3$. Third, the data were analysed using regression using REML as specified in Section 4.4 (the `lmer` () function from the `lmerTest` package in R), omitting the effects of the patient covariates associated with $\delta_2$ and $\delta_3$. Fourth, the following quantities were extracted from the fitted model: the intervention effect $\hat{\delta}_1$, its standard error (SE($\hat{\delta}_1$)), estimates of the variance components $\sigma^2_{u1}$ and $\sigma^2_{v1}$, whether either variance component was estimated to be zero (a boundary estimate), and indicators for false positives (Type I error) and false negatives (Type II error) in testing the null hypothesis that $\delta_1 = 0$ against the alternative hypothesis that $\delta_1 \neq 0$ with a 5% significance level. Fifth, steps 1 to 4 were repeated for 10000 replications, and the average of each estimated quantity was computed over all replications.

The results of our simulation study are summarised in Table 3 below for example 1 (the completely randomised factorial design). The results for examples 2 and 3 can be found in Supplementary File G.

[Insert Table 3 about here]

Allowing for Monte Carlo error, the intervention effect, $\delta_1$, was estimated without bias, regardless of the size of $\delta_2$ and $\delta_3$, across all five randomisation methods. The between-therapist variance, $\sigma^2_{u1}$, was estimated to be about 0.1 when $\delta_2$ and $\delta_3$ were both zero but increased substantially to near 1.0 when $\delta_2 = 0.2$ and therapists were non-randomly allocated to patients ($D_{11}$, $D_{21}$ or $D_{31}$), as expected. When therapists were randomised to patients ($D_{41}$ or $D_{51}$), the between-therapist variance, $\sigma^2_{u1}$, was again estimated to be about 0.1, showing that randomising therapists to patients aids unbiased estimation of the between-therapist variance. Likewise, the between-therapist variance in the intervention effect, $\sigma^2_{v1}$, was estimated to be near 0.15 when $\delta_2$ and $\delta_3$ were both zero. It increased substantially to about 1.0 when $\delta_3 = 0.2$ and therapists were non-randomly allocated to patients *after* the intervention was allocated with knowledge of this randomised intervention ($D_{11}$), as expected. When $\delta_3 = 0$ and therapists were non-randomly allocated to patients *before* allocating the intervention, thus concealing the intervention ($D_{21}$ or $D_{31}$), we constrained the between-therapist variance in the intervention effect, $\sigma^2_{v1}$, to be about 0.15. Yet, when $\delta_3 = 0.2$ and therapists were randomised to patients ($D_{41}$ or $D_{51}$), $\sigma^2_{v1}$ was again estimated be about 0.15. This indicates that concealing the randomised intervention when allocating the therapist, or randomising therapists to patients, facilitates the unbiased estimation of the between-therapist variance in the intervention effect. The consequences of increases in the estimated variance components can be seen in the estimated standard error of $\hat{\delta}_1$ and the Type II error.

The standard error of $\hat{\delta}_1$ was estimated to increase substantially where the between-therapist variance in the intervention effect, $\sigma^2_{v1}$, was estimated to increase substantially (when $\delta_3 = 0.2$ for $D_{11}$). This substantial increase in the standard error of $\hat{\delta}_1$ was associated with a similar substantial increase in the estimated Type II error rate, as expected. The standard error of $\hat{\delta}_1$ was estimated to increase modestly where $\delta_2 = 0.2$ across all five randomisation methods, associated with a similar modest increase in the estimated Type II error rate. The estimated substantial increase in the between-therapist variance, $\sigma^2_{u1}$, observed if therapists were non-randomly assigned to patients ($D_{11}$, $D_{21}$ and $D_{31}$), made little or no difference to the estimated standard error of $\hat{\delta}_1$ or the associated Type II error rate.

The Type I error rate was consistently estimated to be 0.05 (5%) regardless of the size of $\delta_2$ and $\delta_3$, across all five randomisation methods, except where $\delta_3 = 0.2$ and the therapists were randomised to patients *after* interventions were randomised to patients ($D_{51}$). Here, the Type I error increased when the covariate related to $\delta_3$ was included in generating the data after interventions were randomised to patients. This issue appeared in $D_{51}$ but not $D_{11}$. The reason is that the covariate related to $\delta_3$ created an imbalance in the types of patients assigned each intervention. Once intervention assignment was fixed, therapists were then randomly selected in a way that balanced this covariate in each level of $I \wedge T$. Because the interventions were already randomised, this balancing created an imbalance in the number of patients in each level of $I \wedge T$, and in turn across interventions with respect to the covariate. As this imbalance was not adjusted for in the analysis model, it sometimes falsely indicated that the intervention effect was statistically significant, even when this effect was only due to the covariate related to $\delta_3$. Moreover, the imbalance associated with this covariate became more pronounced with larger sample sizes, with greater inflation of the Type I error rate as the sample size increased.

We found that the variance components $\sigma^2_{u1}$ and $\sigma^2_{v1}$ were estimated to be zero in a large proportion of simulations giving boundary estimates. When $\sigma^2_{v1}$ is estimated to be zero, the analysis is equivalent to an analysis based on the assumption that $\sigma^2_{v1}$ is known to be zero [58]. Thus, the degrees of freedom available to test the null hypothesis that $\delta_1 = 0$ increased substantially.

## 6 Discussion

The statistical literature to date has recognised the sample size and analysis implications of treatment-related clustering due to therapists delivering interventions in individually-randomised parallel-group trials. We have used and adapted the Design of Experiments approach described in [42] to formally set out the statistical implications of considering therapists as a random treatment factor in a family of orthogonal factorial designs, and their associated data analyses. We refer to these randomisation designs as ‘completely randomised’, ‘randomised block’ and ‘multicentre randomised block’ factorial designs, respectively, to reflect the absence of any blocking (i.e., stratification) factors, the inclusion of categorical time (Batch) as a blocking factor and the further inclusion of Centre as a blocking factor. We have shown how a formal approach to designing trials of therapist-delivered interventions leads to recommendations about how to randomise and analyse such trials. We have reported the results of a small simulation study, which back up the theory. Our simulations indicate the impact of (i) randomising therapists to patients on estimation of the between-therapist variance, (ii) concealing the intervention in assigning therapists to patients or randomising therapists to patients on estimation of the between-therapist variance in the intervention effect, (iii) consequent impact on the standard error of the intervention effect and its Type II error, and (iv) the impact of separately randomising therapists to patients and randomising interventions to patients on its Type I error.

A challenge we have started to address is how to handle factors that would ideally be randomised but due to practical constraints cannot be. These were Batch and Centre here. It leads to designs in which the treatment structure is not fully representative of the complex intervention design. In this paper, we have suggested blocking these factors to ensure that factors that affect the delivery of the intervention package are orthogonal to the treatment factors, enabling unbiased estimation of treatment effects and better estimation of interactions between the components of an intervention package. Further work is needed to inform decisions about which factors to include in both the randomisation and the analysis, and which to include just in the analysis, where there are too many to include all in the randomisation. We have included all the block-treatment interactions in the primary analysis. Further work is needed to better appreciate the importance of these interactions in routine data and in historical trials.

In our simulation study, blocking the randomisation of interventions to patients by therapists made no discernible difference beyond the necessity to assign therapists to patients beforehand and therefore to conceal the intervention when allocating the therapist. This implies that some lack of balance in the data is unlikely to have much of an effect. However, randomising therapist-intervention combinations when compared to randomising the intervention and therapist to patients separately had a clear effect on the Type I error. This implies that our proposed approach to randomisation is preferable. If one of the secondary aims of the trial is to unbiasedly estimate the between-therapist variance and between-therapist variance in the intervention effect, we have shown that either randomisation method is to be preferred. At a minimum, we suggest concealing the intervention when assigning therapists to patients to avoid inflating the Type II error and decreasing the power to detect the intervention effect. Further simulation work is needed to fully evaluate performance in a range of realistic scenarios.

Based on the statistical considerations, we would recommend randomising therapists to patients (or at least concealing the intervention when assigning therapists to patients) but the barriers and enablers to making this feasible in practice need to be understood. The Second Sheffield Psychotherapy Project [46] employed research therapists to deliver the interventions, which gave the researchers control over their randomisation. Many more recent trials (including PRaCTICED [44]) use therapists embedded in routine clinical services, giving less control to the researchers. Building on [6], work is planned to build (i) variation in therapist capacity to take on patients, (ii) there only being a subset of therapists in a centre available to take on a patient at the point of randomisation, and (iii) therapist turnover, into a more flexible randomisation method. Moreover, the number of centres or batches may be limited in practice, questioning the assumption we have made that Batch and Centre contribute random effects. Further work is planned to understand the sample size implications of the randomisation designs that we have proposed here using an optimal design approach.

The randomisation designs proposed here are specifically for the case where each therapist delivers each intervention (the crossed therapist-intervention design [6]), there is a single therapist-per-patient and a single continuous outcome. Other therapist-intervention designs lead to randomisation designs with potentially different statistical and practical considerations. Planned and unplanned multiple therapist-per-patient designs are common, as are longitudinal outcomes and other outcome types. We did not consider the possibility of intervention contamination, nor the possibility of therapists taking on patients in continuous time. These are areas that require further work. Finally, we assumed that the therapists are a random sample from a population of interest. This may not be the case in practice. If a convenience sample of therapists is used, caution must be given when considering the generalisability of the trial results.

To our knowledge, this is the first time that a formal Design of Experiments approach has been taken to designing trials of therapist-delivered interventions. This has led to a better understanding of the data analysis, which follows the design. Based on this, we would recommend that: (i) the primary data analysis should fully reflect the randomisation, including blocking by time; (ii) Satterthwaite's [57] method for calculating the degrees of freedom should be used routinely; (iii) treatment contrasts are orthogonal, estimating main effects and interactions; (iv) although many of the calculations can be done by hand, software is needed enabling ANOVA to be used more widely if there are 'dependent' random effects. It is possible that use of ANOVA for the data analysis of these trials would lead to fewer unintended consequences of boundary estimates on the degrees of freedom when testing the intervention effect. This needs further research.

More purposeful trial design has the potential to lead to better evidence on a range of complex interventions. We have focused here on psychotherapy, but the methods described are applicable to trials of physiotherapy, rehabilitation, surgery and any complex intervention delivered by a person.

**Figure 1: Complex intervention designs**

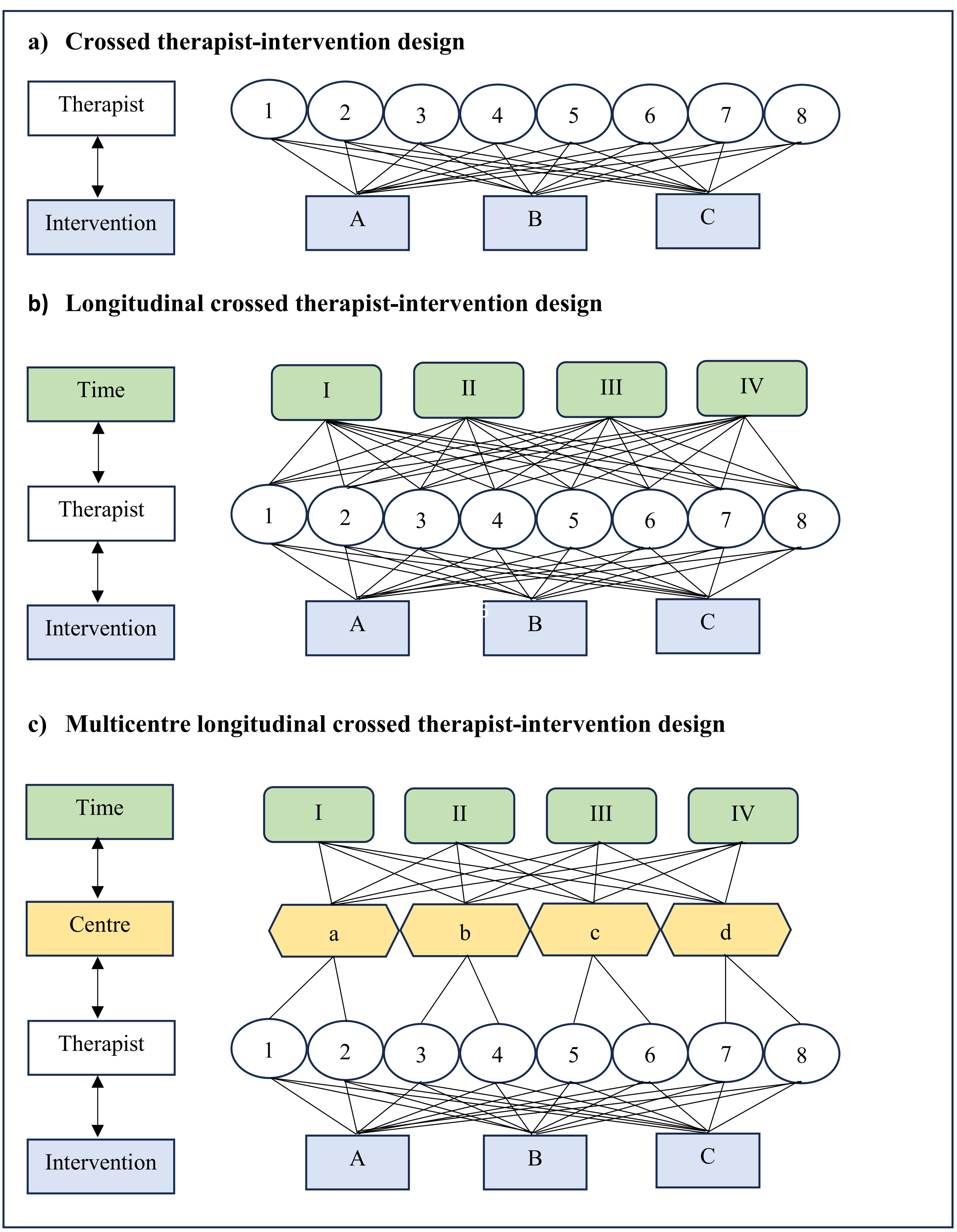

**Figure 2: Randomisation diagrams**

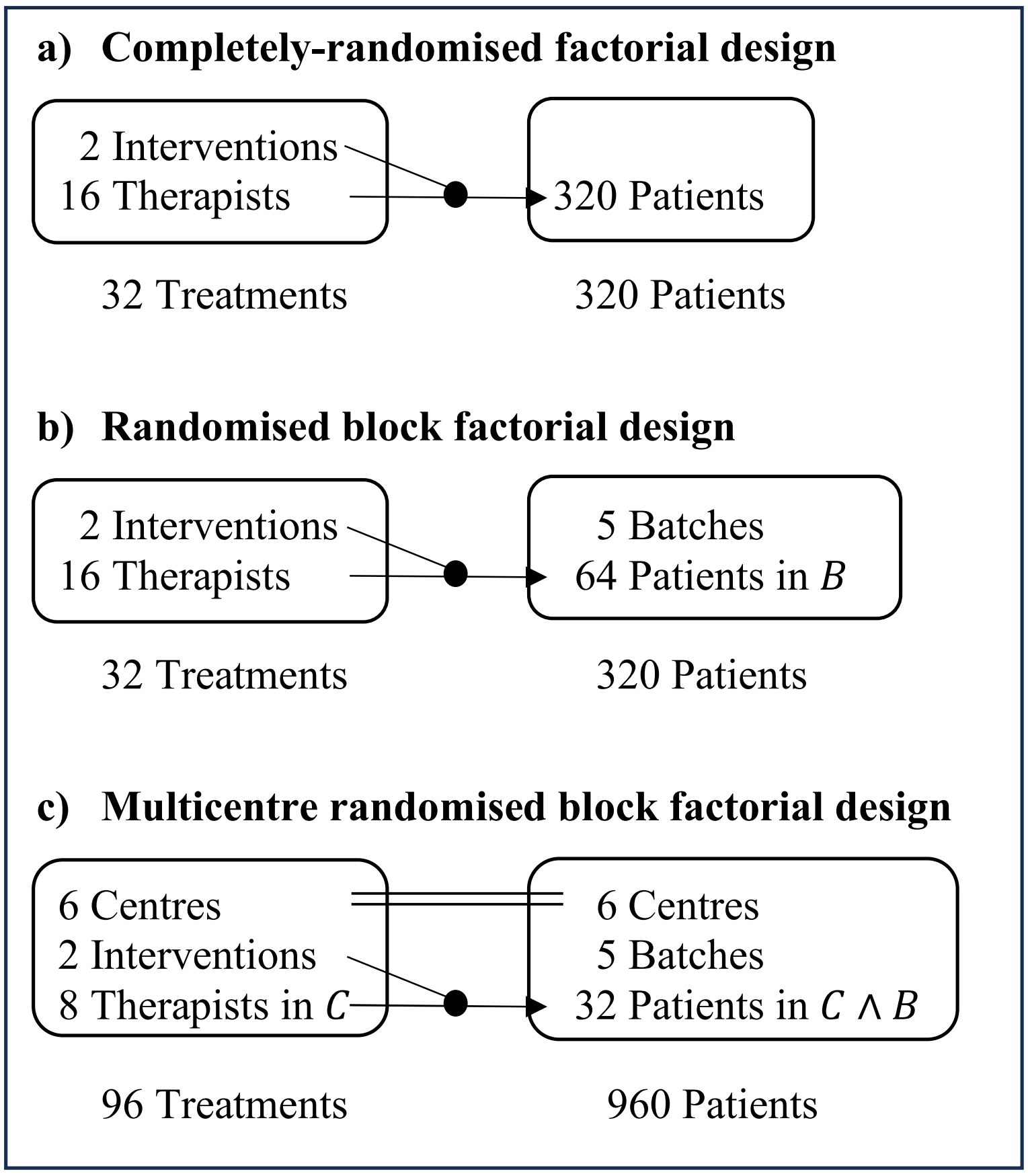

**Figure 3: Hasse diagrams for the random and fixed structures.**

**a) Completely-randomised factorial design**

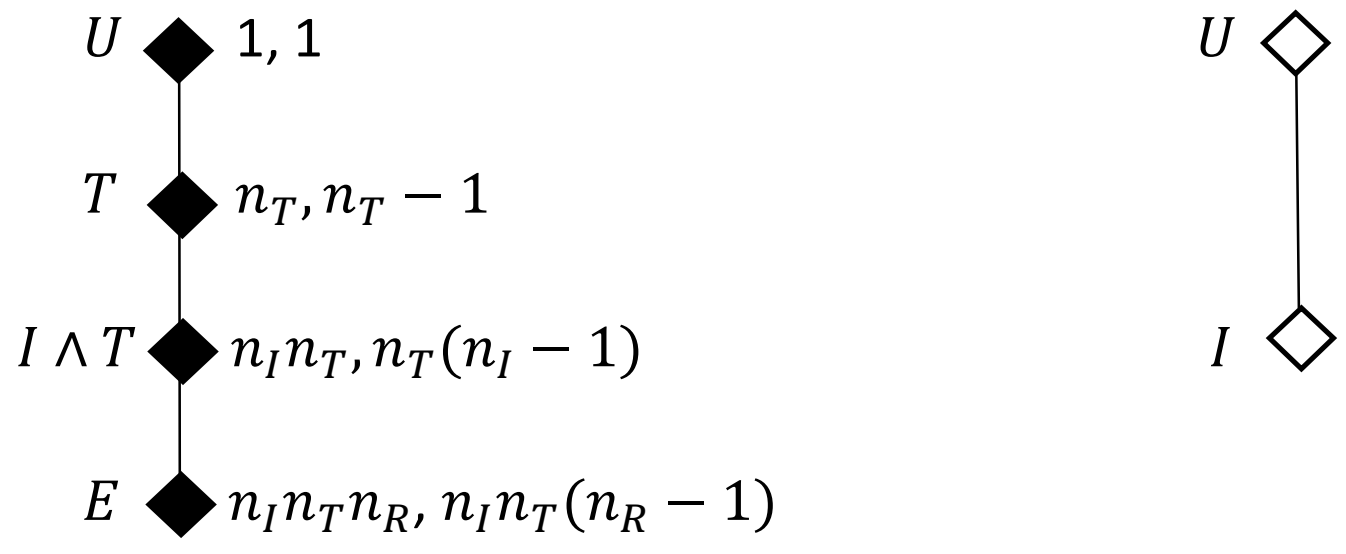

$U$ 1, 1

$I$ $n_I, n_I - 1$

(i) Random structure

(ii) Fixed structure

**b) Randomised block factorial design**

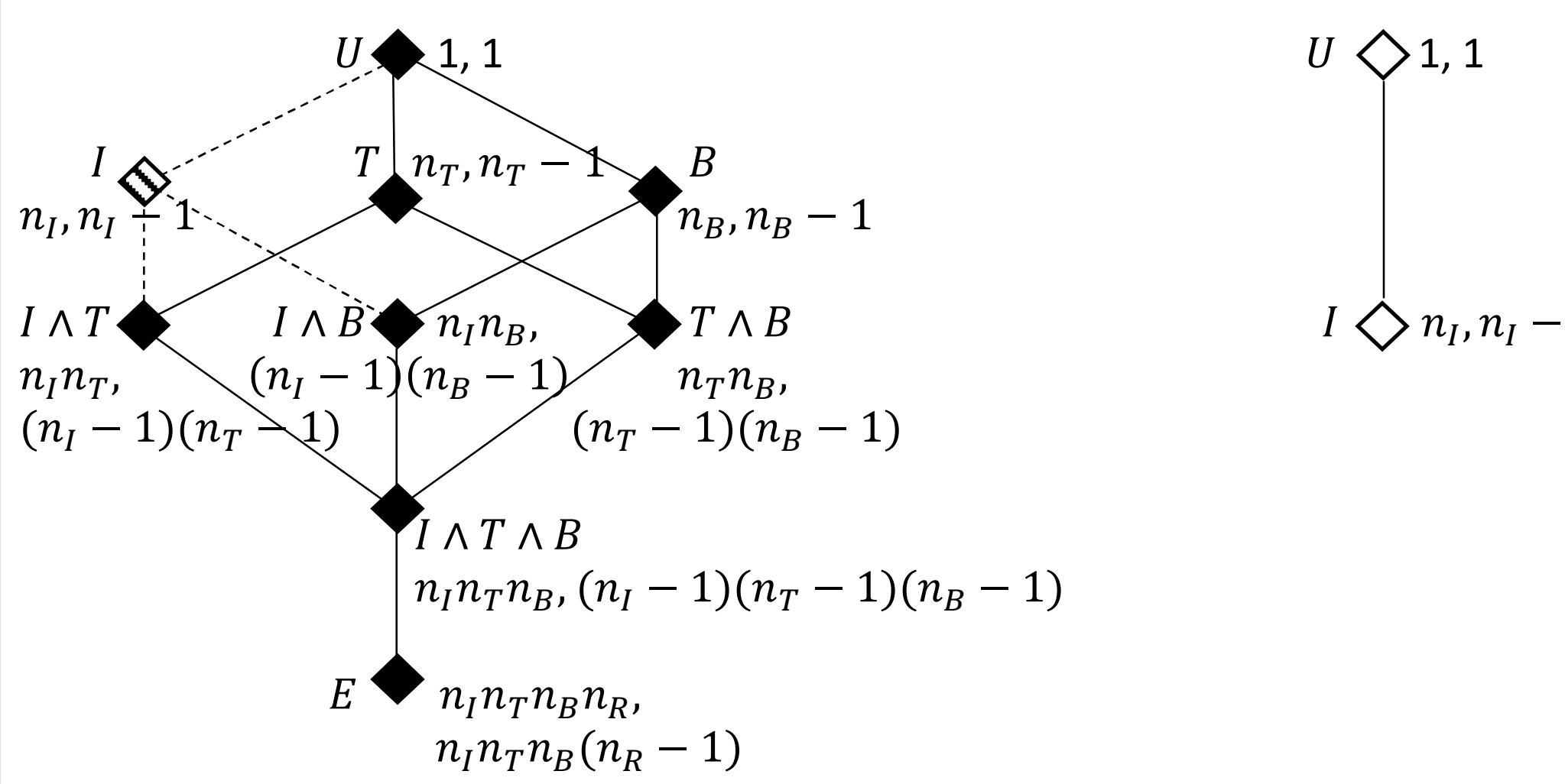

$U$ 1, 1

$I$ $n_I, n_I - 1$

(i) Random structure

(ii) Fixed structure

**c) Multicentre randomised block factorial design**

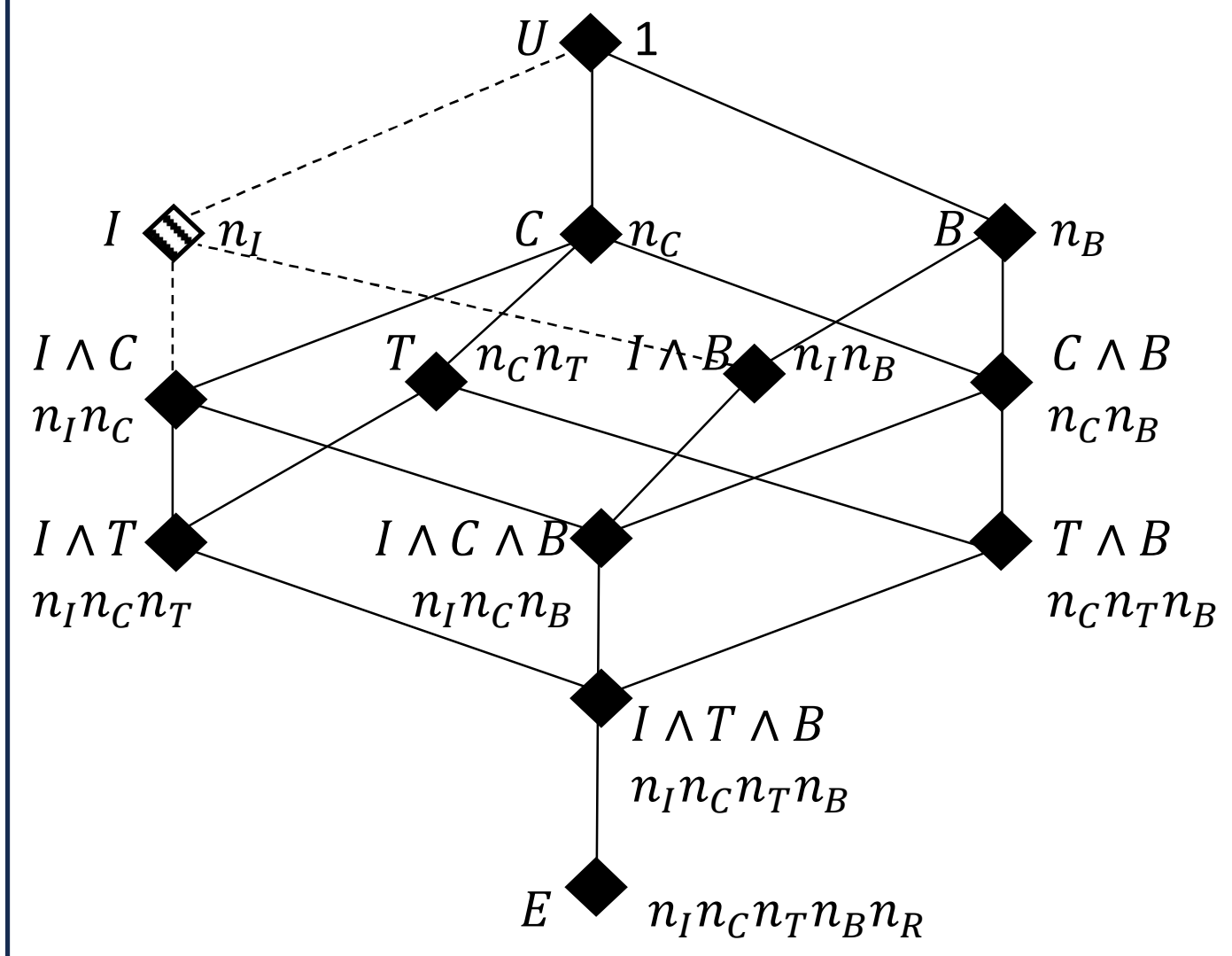

$U$ 1

$I$ $n_I$

(i) Random structure

(ii) Fixed structure

**Table 1. Analysis of variance (ANOVA).**

| Stratum | Source | Degrees of Freedom | Expected Mean Squares |
|---|---|---|---|
| **a) Completely randomised factorial design** | | | |
| $W_0$ Mean | Mean | 1 | $\lVert\boldsymbol{\tau}_0\rVert^2+\xi_0$ |
| $W_T$ Therapists | Therapists | 15 | $\xi_T$ |
| $W_{I\wedge T}$ $I \wedge T$ Treatments | Interventions | 1 | $\lVert\boldsymbol{\tau}_I\rVert^2+\xi_{I\wedge T}$ |
| | Residual | 15 | $\xi_{I\wedge T}$ |
| | Total | 16 | |
| $W_E$ Patients | Patients | 288 | $\xi_E$ |
| Total | | 320 | |
| | | | |
| **b) Randomised block factorial design** | | | |
| $W_0$ Mean | Mean | 1 | $\lVert\boldsymbol{\tau}_0\rVert^2+\xi_0$ |
| $W_I$ Interventions | Interventions | 1 | $\lVert\boldsymbol{\tau}_I\rVert^2+\xi_{I\wedge T}+\xi_{I\wedge B}-\xi_{I\wedge T\wedge B}$ |
| | Total | 1 | |
| $W_T$ Therapists | Therapists | 15 | $\xi_T$ |
| $W_B$ Batches | Batches | 4 | $\xi_B$ |
| $W_{I\wedge T}$ $I \wedge T$ | $I \wedge T$ | 15 | $\xi_{I\wedge T}$ |
| $W_{I\wedge B}$ $I \wedge B$ | $I \wedge B$ | 4 | $\xi_{I\wedge B}$ |
| $W_{T\wedge B}$ $T \wedge B$ | $T \wedge B$ | 60 | $\xi_{T\wedge B}$ |
| $W_{I\wedge T\wedge B}$ $I \wedge T \wedge B$ | $I \wedge T \wedge B$ | 60 | $\xi_{I\wedge T\wedge B}$ |
| $W_E$ Patients | Patients | 160 | $\xi_E$ |
| Total | | 320 | |
| | | | |
| **c) Multicentre randomised block factorial design** | | | |
| $W_0$ Mean | Mean | 1 | $\lVert\boldsymbol{\tau}_0\rVert^2+\xi_0$ |
| $W_I$ Interventions | Interventions | 1 | $\lVert\boldsymbol{\tau}_I\rVert^2+\xi_{I\wedge C}+\xi_{I\wedge B}-\xi_{I\wedge C\wedge B}$ |
| | Total | 1 | |
| $W_C$ Centres | Centres | 5 | $\xi_C$ |
| $W_B$ Batches | Batches | 4 | $\xi_B$ |
| $W_{I\wedge C}$ | $I \wedge C$ | 5 | $\xi_{I\wedge C}$ |
| $W_T$ Therapists | Therapists | 42 | $\xi_T$ |
| $W_{I\wedge B}$ | $I \wedge B$ | 4 | $\xi_{I\wedge B}$ |
| $W_{C\wedge B}$ | $C \wedge B$ | 20 | $\xi_{C\wedge B}$ |
| $W_{I\wedge T}$ | $I \wedge T$ | 42 | $\xi_{I\wedge T}$ |
| $W_{I\wedge C\wedge B}$ | $I \wedge C \wedge B$ | 20 | $\xi_{I\wedge C\wedge B}$ |
| $W_{T\wedge B}$ | $T \wedge B$ | 168 | $\xi_{T\wedge B}$ |
| $W_{I\wedge T\wedge B}$ | $I \wedge T \wedge B$ | 168 | $\xi_{I\wedge T\wedge B}$ |
| $W_E$ Patients | Patients | 480 | $\xi_E$ |
| Total | | 960 | |

**Table 2. Data generating mechanism.**

| Model Parameter | Example | | |
|---|---|---|---|
| | a) 1 Centre, 1 Batch (320 Patients) | b) 1 Centre, 5 Batches (320 Patients) | c) 6 Centres, 5 Batches (960 Patients) |
| $\delta_0$ (Mean) | 0.000 | 0.000 | 0.000 |
| $\delta_1$ $(I)$ | 0.315 | 0.180 | 0.180 |
| $\sigma_{u1}^2$ $(T)$ | 0.100 | 0.100 | 0.100 |
| $\sigma_{v1}^2$ $(I \wedge T)$ | 0.150 | 0.150 | 0.100 |
| $\sigma_e^2$ $(E)$ | 0.750 | 0.350 | 0.200 |
| $\sigma_{u2}^2$ $(B)$ | - | 0.100 | 0.100 |
| $\sigma_{u3}^2$ $(T \wedge B)$ | - | 0.100 | 0.050 |
| $\sigma_{v2}^2$ $(I \wedge B)$ | - | 0.100 | 0.100 |
| $\sigma_{v3}^2$ $(I \wedge T \wedge B)$ | - | 0.100 | 0.050 |
| $\sigma_{u4}^2$ $(C)$ | - | - | 0.100 |
| $\sigma_{u5}^2$ $(C \wedge B)$ | - | - | 0.050 |
| $\sigma_{v4}^2$ $(I \wedge C)$ | - | - | 0.100 |
| $\sigma_{v5}^2$ $(I \wedge C \wedge B)$ | - | - | 0.050 |

**Table 3. Simulation results for the completely randomised factorial design.**

| $\boldsymbol{\delta_3}$ | $\boldsymbol{\delta_2}$ | $\boldsymbol{\hat{\delta}_1}$ | | **SE($\boldsymbol{\hat{\delta}_1}$)** | | $\boldsymbol{\hat{\sigma}^2_{u1}}$ | | $\boldsymbol{\hat{\sigma}^2_{v1}}$ | | **Type I Error** | | **Type II Error** | | **Boundary Estimates** | |
|---|---|---|---|---|---|---|---|---|---|---|---|---|---|---|---|
| 1) Comparison of assigning therapists to patients *after* ($D_{11}$) and *before* ($D_{21}$) randomising interventions to patients | | | | | | | | | | | | | | | |
| | | $\boldsymbol{D_{11}}$ | $\boldsymbol{D_{21}}$ | $\boldsymbol{D_{11}}$ | $\boldsymbol{D_{21}}$ | $\boldsymbol{D_{11}}$ | $\boldsymbol{D_{21}}$ | $\boldsymbol{D_{11}}$ | $\boldsymbol{D_{21}}$ | $\boldsymbol{D_{11}}$ | $\boldsymbol{D_{21}}$ | $\boldsymbol{D_{11}}$ | $\boldsymbol{D_{21}}$ | $\boldsymbol{D_{11}}$ | $\boldsymbol{D_{21}}$ |
| 0 | 0 | 0.315 | 0.315 | 0.083 | 0.083 | 0.08 | 0.08 | 0.14 | 0.13 | 0.05 | 0.05 | 0.06 | 0.06 | 14% | 15% |
| 0 | 0.2 | 0.314 | 0.314 | 0.101 | 0.101 | 1.05 | 1.04 | 0.12 | 0.12 | 0.05 | 0.05 | 0.18 | 0.18 | 9% | 9% |
| 0.2 | 0 | 0.315 | - | 0.194 | - | 1.10 | - | 1.00 | - | 0.05 | - | 0.66 | - | 9% | - |
| 0.2 | 0.2 | 0.315 | - | 0.208 | - | 0.95 | - | 1.08 | - | 0.05 | - | 0.72 | - | 17% | - |
| 2) Comparison of assigning therapists to patients *before* ($D_{21}$) randomising interventions to patients with blocking the randomisation of interventions to patients by therapists ($D_{31}$) | | | | | | | | | | | | | | | |
| | | $\boldsymbol{D_{21}}$ | $\boldsymbol{D_{31}}$ | $\boldsymbol{D_{21}}$ | $\boldsymbol{D_{31}}$ | $\boldsymbol{D_{21}}$ | $\boldsymbol{D_{31}}$ | $\boldsymbol{D_{21}}$ | $\boldsymbol{D_{31}}$ | $\boldsymbol{D_{21}}$ | $\boldsymbol{D_{31}}$ | $\boldsymbol{D_{21}}$ | $\boldsymbol{D_{31}}$ | $\boldsymbol{D_{21}}$ | $\boldsymbol{D_{31}}$ |
| 0 | 0 | 0.315 | 0.313 | 0.083 | 0.082 | 0.08 | 0.08 | 0.13 | 0.13 | 0.05 | 0.05 | 0.06 | 0.06 | 15% | 14% |
| 0 | 0.2 | 0.314 | 0.315 | 0.101 | 0.100 | 1.04 | 1.05 | 0.12 | 0.12 | 0.05 | 0.05 | 0.18 | 0.17 | 9% | 9% |
| 0.2 | 0 | - | - | - | - | - | - | - | - | - | - | - | - | - | - |
| 0.2 | 0.2 | - | - | - | - | - | - | - | - | - | - | - | - | - | - |
| 3) Comparison of blocking the randomisation of interventions to patients by therapists ($D_{31}$) and randomising the therapist-intervention combination to patients (our proposed method) ($D_{41}$) | | | | | | | | | | | | | | | |
| | | $\boldsymbol{D_{31}}$ | $\boldsymbol{D_{41}}$ | $\boldsymbol{D_{31}}$ | $\boldsymbol{D_{41}}$ | $\boldsymbol{D_{31}}$ | $\boldsymbol{D_{41}}$ | $\boldsymbol{D_{31}}$ | $\boldsymbol{D_{41}}$ | $\boldsymbol{D_{31}}$ | $\boldsymbol{D_{41}}$ | $\boldsymbol{D_{31}}$ | $\boldsymbol{D_{41}}$ | $\boldsymbol{D_{31}}$ | $\boldsymbol{D_{41}}$ |
| 0 | 0 | 0.313 | 0.315 | 0.082 | 0.082 | 0.08 | 0.08 | 0.13 | 0.13 | 0.05 | 0.05 | 0.06 | 0.06 | 14% | 14% |
| 0 | 0.2 | 0.315 | 0.313 | 0.100 | 0.112 | 1.05 | 0.07 | 0.12 | 0.10 | 0.05 | 0.05 | 0.17 | 0.25 | 9% | 40% |
| 0.2 | 0 | - | - | - | - | - | - | - | - | - | - | - | - | - | - |
| 0.2 | 0.2 | - | - | - | - | - | - | - | - | - | - | - | - | - | - |
| 4) Comparison of randomising the therapist-intervention combination to patients (our proposed method) ($D_{41}$) and randomising interventions to patients, then randomising therapists to patients ($D_{51}$) | | | | | | | | | | | | | | | |
| | | $\boldsymbol{D_{41}}$ | $\boldsymbol{D_{51}}$ | $\boldsymbol{D_{41}}$ | $\boldsymbol{D_{51}}$ | $\boldsymbol{D_{41}}$ | $\boldsymbol{D_{51}}$ | $\boldsymbol{D_{41}}$ | $\boldsymbol{D_{51}}$ | $\boldsymbol{D_{41}}$ | $\boldsymbol{D_{51}}$ | $\boldsymbol{D_{41}}$ | $\boldsymbol{D_{51}}$ | $\boldsymbol{D_{41}}$ | $\boldsymbol{D_{51}}$ |
| 0 | 0 | 0.315 | 0.315 | 0.082 | 0.082 | 0.08 | 0.08 | 0.13 | 0.13 | 0.05 | 0.05 | 0.06 | 0.06 | 14% | 14% |
| 0 | 0.2 | 0.313 | 0.315 | 0.112 | 0.111 | 0.07 | 0.07 | 0.10 | 0.10 | 0.05 | 0.05 | 0.25 | 0.36 | 40% | 39% |
| 0.2 | 0 | - | 0.312 | - | 0.112 | - | 0.08 | - | 0.10 | - | 0.27 | - | 0.35 | | 37% |
| 0.2 | 0.2 | - | 0.316 | - | 0.137 | - | 0.07 | - | 0.09 | - | 0.20 | - | 0.43 | | 57% |

Note: $\hat{\delta}_1$ is the fixed intervention effect, SE($\hat{\delta}_1$) is the standard error of $\hat{\delta}_1$, $\hat{\sigma}^2_{u1}$ is the between-therapist variance, $\hat{\sigma}^2_{v1}$ is the between-therapist variance in the intervention effect, $\delta_2$ is the fixed coefficient of the baseline patient covariate associated with how therapists are allocated to patients and $\delta_3$ is the fixed coefficient of baseline patient covariate associated with how therapists are allocated to patients with knowledge of the randomised intervention. Boundary estimates are where $\sigma^2_{u1}$ or $\sigma^2_{v1}$ is estimated to be zero and a warning is given by the software. The model converged in every case.